\documentclass[%
 aip,
 amsmath,amssymb,
 reprint,%
]{revtex4-2}

\usepackage{graphicx}
\usepackage{dcolumn}
\usepackage{bm}

\usepackage[utf8]{inputenc}
\usepackage[T1]{fontenc}
\usepackage{mathptmx}

\usepackage{braket}
\usepackage{amsmath,amssymb}
\usepackage{physics}
\usepackage{amsmath}
\usepackage{bbold}
\usepackage{braket}
\usepackage{xcolor}
\usepackage{placeins}
\usepackage{ulem}
\usepackage{natbib}

\begin{document}


\title[]{Non-Linear Correlation Functions 
and Zero-Point Energy Flow in Mixed Quantum-Classical Semiclassical Dynamics}

\author{Shreyas Malpathak}
\author{Nandini Ananth}%
 \email{ananth@cornell.edu}
\affiliation{Department of Chemistry and Chemical Biology, Baker Laboratory, Cornell University
Ithaca,14853 NY, USA}%


\date{\today}

\begin{abstract}
Mixed Quantum Classical (MQC)-IVR is a recently introduced 
semiclassical framework that allows for selective 
quantization of the modes of a complex system. 
In the quantum limit, MQC reproduces the semiclassical Double Herman-Kluk IVR results,
accurately capturing nuclear quantum coherences 
and conserving zero-point energy. However, in the classical limit,
while MQC mimics the Husimi-IVR for real-time correlation functions 
with linear operators, it is significantly
less accurate for non-linear correlation functions with 
errors even at time zero. Here, we identify the origin of this discrepancy in
the MQC formulation and propose a modification. We analytically show that the 
modified MQC approach is exact for all correlation functions at time zero,
and in a study of zero-point energy (ZPE) flow, we numerically demonstrate 
that it correctly obtains the quantum and classical limits as a function of time. 
Interestingly, while classical-limit MQC simulations show the expected, 
unphysical ZPE leakage, we find it is possible to predict and 
even modify the direction of ZPE flow through selective 
quantization of the system, with the quantum-limit modes accepting energy additions 
but preserving the minimum quantum mechanically required energy.
\end{abstract}

\maketitle

\section{Introduction}
Probing the effects of the quantum nature of nuclei and
nuclear-electronic coupling in chemical and biological 
systems has been the focus of extensive research in recent 
years.~\cite{Cao2020,Markland2018,Ananth2022} 
Numerically exact methods to simulate quantum dynamics 
have found applications in chemical systems 
of modest sizes~\cite{Schulze2016,Lyu2022} 
but their expensive scaling with system size 
remains a challenge. Path integral based methods 
that rely on classical trajectories, such as 
Matsubara dynamics,~\cite{Hele2015b,Hele2015a} 
Ring Polymer Molecular Dynamics (RPMD),~\cite{Craig2004,Craig2005} 
Centroid Molecular Dynamics (CMD),~\cite{Cao1994c,Jang1999} and 
Multistate Ring Polymer Molecular Dynamics~\cite{Ananth2013,Duke2016}
have significantly more favorable scaling laws, 
but fail to capture nuclear quantum coherences.

Semiclassical (SC) Initial Value Representation (IVR) methods have emerged as a rigorous alternative for the simulation of 
quantum processes using near-classical trajectories. 
A hierarchy of SC approximations have been established
with the most accurate, quantum-limit SC methods able to describe
quantum effects such as zero point energy, tunneling, nonadiabatic 
and interference effects.~\cite{Miller2001a,Miller2009} 
Unfortunately, these quantum-limit methods require the evaluation
of a complex oscillatory integrand and the resulting 
numerical sign problem limits applications to 
high-dimensional systems.
Classical-limit SC 
methods, like Linearized SC-IVR,~\cite{Wang1998,Shi2003}
do not capture interference effects but employ 
only classical trajectories making them suitable 
for the simulation of condensed phase processes 
where quantum coherences are 
short-lived.~\cite{Liu2015,Poulsen2005b,Sun2018}

The recently introduced mixed quantum-classical SC method, 
MQC-IVR,~\cite{Antipov2015,Church2017,Malpathak2022} has 
been shown to reduce the cost of quantum-limit SC correlation 
function calculations using a modified Filinov filtration 
technique~\cite{Filinov1986a,Makri1987c,Makri1988b} 
to mitigate the effects of oscillatory phase. 
In the limit of small Filinov filter for all degrees 
of freedom (dofs) the MQC correlation function 
becomes identical to a quantum-limit SC-IVR correlation function, 
specifically the Double Herman-Kluk IVR (DHK-IVR).~\cite{Herman1984,Kay1994a,Herman1997,Thoss2001a}
In the limit of large parameters, for linear operators, MQC
corresponds to a classical-limit, linearized method, specifically the 
Husimi-IVR.~\cite{Herman1999,Miller2001a,Thoss2001a,YiZhao2002,Wright2004c,LohoChoudhury2020} 
As the name suggests, the MQC framework uniquely offers a 
path to selective quantization \textemdash\, by filtering the phase 
contribution from different dofs to different extents, it is 
possible to treat some modes in the quantum limit and 
others in the classical limit.

In previous work, for a range of 1D and 2D model systems,
MQC has been numerically shown to gradually tune linear 
correlation functions from the quantum to the 
classical limit,~\cite{Antipov2015,Church2017}
and to capture nuclear coherence effects
in nonadiabatic scattering models.~\cite{Church2018}
The analytic mixed quantum-classical IVR (AMQC-IVR) 
method was introduced more recently to treat a 
handful of system dofs in the quantum-limit 
while treating the rest in the classical-limit, 
enabling the calculation of thermal
reaction rates in high-dimensional system-bath models.~\cite{Church2019a}
Despite these successes, a recent study showed that 
for correlation functions of operators that are not linear in 
position and momentum, MQC 
is less successful particularly in the classical limit
where it yields inaccurate values even at time zero.~\cite{LohoChoudhury2020} 
Given that both DHK-IVR and Husimi-IVR are, in general,
exact for correlation functions at time zero, 
this is a rather startling observation.

In this paper, we demonstrate the origin of the 
problem: for operators that are not linear in 
position and momentum, the classical-limit
MQC correlation function does not coincide with 
the corresponding Husimi-IVR expression. We propose
a general strategy to modify MQC such that
its classical limit reproduces the Husimi-IVR 
expression for both linear and non-linear
correlation functions.
We then use this modified MQC expression to characterize
zero-point energy (ZPE) flow in a series of model systems.

Quasi-classical and linearized SC 
methods typically exhibit ZPE `leakage' in time,
with high frequency modes losing zero-point energy 
to lower frequency ones.\cite{Bowman1989,Miller1989,Habershon2009} 
This non-conservation of ZPE 
is attributed to the use of classical trajectories 
without associated phase terms that are necessary 
to capture interference effects, 
and over the years, several mitigation strategies
have been proposed.~\cite{Lu1988,Bowman1989,
Miller1989,Nyman1990,Varandas1992,Sewell1992,Alimi1992,Peslherbe1994,Lim1995,
Schlier1995,McCormack1995,Guo1996,Stock1999,Xie2006,Brieuc2016} 
In contrast, quantum-limit SC methods like DHK-IVR have been 
shown to conserve ZPE in model systems.\cite{Buchholz2018} 
Here, we investigate ZPE flow in MQC simulations where the strength 
of Filinov filter is varied from the quantum limit to the classical 
limit. We then demonstrate strategies to control
and indeed direct ZPE flow in the mixed quantum-classical 
limit using a coupled oscillator model systems
constructed to mimic the network of 
connections observed in realistic systems. 

The paper is organized as follows: Sec.~\ref{sec:theory} 
begins with a brief review of the Wigner and Husimi 
phase space formulations of quantum mechanics, 
inspects the MQC expression for non-linear 
correlation functions in the classical limit, 
and suggests a general strategy to modify MQC.
Sec.~\ref{sec:simulation} describes the model coupled 
oscillator systems studied here, Sec.~\ref{sec:results}
demonstrates how ZPE flow in these models can be modified
by tuning the extent of quantization of the individual
oscillators, and Sec.~\ref{sec:conclusion} concludes.

\section{Theory} \label{sec:theory}
\subsection{Phase-space formulation of Quantum Mechanics}
In the phase space formulation of quantum mechanics,~\cite{Cohen1966,Lee1995} 
expectation values of operators are calculated 
analogous to classical mechanics,
\begin{align}
    \langle\hat{B}\rangle = \int d \textbf{z} \rho(\textbf{z})B(\textbf{z}),   
\end{align}
where $\textbf{z} = (p,q)$ is the phase space point, and the functional form of the phase space density $\rho(\textbf{z})$ 
and the function $B(\textbf{z})$ corresponding to $\hat{B}$ vary 
based on the formulation. 
Using Cohen's unified classification framework,~\cite{Cohen1966}
the phase space density is expressed as,
\begin{align}
    \rho(\textbf{z}) & = \frac{1}{4\pi^2}\int_{-\infty}^{\infty} 
    d\zeta \int_{-\infty}^{\infty} d\eta \, f\left(\xi,\eta \right) \notag \\
    & \times 
    \text{ Tr}\left[ e^{i\zeta(\hat{x}-x)+i\eta(\hat{p}-p)}\hat{\rho} \right],
    \label{phase-density}
\end{align}
and the phase space function of $\hat B$ as, 
\begin{align}
  B(\textbf{z}) & = \frac{2\pi\hbar}{4\pi^2}\int_{-\infty}^{\infty} d\zeta \int_{-\infty}^{\infty} 
  d\eta \, f^{-1}\left(\xi,\eta \right)  \notag \\
  & \times \text{ Tr}\left[ e^{i\zeta(\hat{x}-x)+i\eta(\hat{p}-p)}\hat{B} \right].
    \label{phase-b}
\end{align}
The function $f(\xi,\eta)$ in Eq.~\ref{phase-b} takes different 
forms in the Wigner\cite{Wigner1932} and Husimi\cite{Husimi1940} 
phase space formulation of quantum mechanics.

In the Wigner formulation, 
\begin{align}
    f\left(\xi,\eta \right) = 1,  
\end{align}
resulting in, 
\begin{align}
    \rho_W(\textbf{z}) = \frac{1}{2\pi\hbar}
    \int d\Delta \mel{q+\frac{\Delta}{2}}{\hat{\rho}}{q-\frac{\Delta}{2}}e^{-ip\Delta/\hbar},
    \label{eq:rhowig}
\end{align}
and, 
\begin{align}
    B_W(\textbf{z}) = \int d\Delta \mel{q+\frac{\Delta}{2}}{\hat{B}}{q-\frac{\Delta}{2}}
    e^{-ip\Delta/\hbar}.
    \label{eq:bwig}
\end{align}
Comparing Eq.~\ref{eq:rhowig} and Eq.~\ref{eq:bwig}, we see 
that the Wigner correspondence rules for the density operator 
and for general operators $\hat{B}$ differ only by a factor 
of $(2\pi\hbar)^{-1}$ that ensures the phase space density is normalized. 

In the Husimi formulation, the correspondence rules are 
different with, 
\begin{align}
f\left(\xi,\eta \right) = e^{-\frac{\zeta^2}{4\gamma}-
\frac{\hbar^2\gamma}{4}\eta^2},  
\end{align}
resulting in,
\begin{align}
    \rho_H(\textbf{z}) = \frac{1}{2\pi\hbar}\mel{\textbf{z}}{\hat{\rho}}{\textbf{z}}, \label{rho-hus}
\end{align}
and, 
\begin{align}
    \Tilde{B}_H(\textbf{z}) &= \frac{2\pi\hbar}{4\pi^2}\int_{-\infty}^{\infty} d\zeta \int_{-\infty}^{\infty} d\eta \, e^{\frac{\zeta^2}{4\gamma}+\frac{\hbar^2\gamma}{4}\eta^2} \notag \\
    & \times \text{ Tr}\left[e^{i\zeta(\hat{x}-x)+i\eta(\hat{p}-p)}\hat{\rho} \right] \notag \\
    &\neq \mel{\textbf{z}}{\hat{B}}{\textbf{z}},
    \label{b-husimi}
\end{align}
referred to as the \textit{anti-Husimi} transform of $\hat{B}$. 
Here, $\ket{\textbf{z}}$ is a coherent state with width $\gamma$ centered at $\textbf{z}$,
\begin{align}
    \braket{x}{\textbf{z}}=\left(\frac{\gamma}{\pi}\right)^{1/4}e^{-\frac{\gamma}{2}(x-q)^2+ip(x-q)/\hbar}.
    \label{cs-def}
\end{align}
The Husimi density in Eq.~\ref{rho-hus} is 
proportional to the diagonal coherent 
state matrix element of the density operator, whereas this is
not true for a general operator $\hat{B}$, 
except in special cases like $\hat{B} \equiv \hat{x} \text{ or } \hat{p}$. 
It is this difference that changes the accuracy of the 
the classical-limit of the MQC correlation function 
even at time zero, as shown in ~\ref{subsec:cft0}.

To find the anti-Husimi transform of an operator,
it is useful to establish a connection between
the Wigner and Husimi functions,~\cite{Harriman1993}
\begin{align}
    B_{H}(\textbf{z}) = \hat{\mathcal{G}}\left(\gamma;q\right)
    \hat{\mathcal{G}}\left(\gamma^{-1}\hbar^{-2};p\right)B_W(\textbf{z}),
    \label{hus-def}
\end{align}
\begin{align}
    \Tilde{B}_H(\textbf{z}) = \hat{\mathcal{G}}^{-1}\left(\gamma;q\right)
    \hat{\mathcal{G}}^{-1}\left(\gamma^{-1}\hbar^{-2};p\right)B_W(\textbf{z}),
    \label{anti-hus-def}
\end{align}
and,
\begin{align}
        B_H(\textbf{z}) = \hat{\mathcal{G}}\left(\frac{\gamma}{2};q\right)\hat{\mathcal{G}}
    \left(\frac{1}{2}\gamma^{-1}\hbar^{-2};p\right)\Tilde{B}_H(\textbf{z}),
    \label{hus-anti-hus}
\end{align}
where $\hat{\mathcal{G}}$ is a Gaussian convolution operator defined as,
\begin{align}
    \hat{\mathcal{G}}(\gamma;x)f(x) &\equiv \left(\frac{\gamma}{\pi}\right)^{1/2}\int dx^{\prime} e^{-\gamma (x-x^{\prime})^2} f(x^{\prime}) \label{gc-1} \\
    & = e^{\frac{1}{4\gamma}\frac{d^2}{dx^2}} f(x) \label{gc-2}.
\end{align}

\subsection{MQC Correlation Functions at Time Zero
}\label{subsec:cft0}

A general, real-time quantum correlation function can be expressed as,
\begin{align}
    C_{AB}(t) = \text{Tr}\left[\hat{\rho}\hat{A}\hat{B}(t)\right].
    \label{cab}
\end{align}
where $\hat H$ is the system Hamiltonian, $\hat{\rho}$ is the density, 
and $\hat A$ and $\hat B$ are the operators evaluated at time zero
and time $t$ respectively. Evaluated in a phase space formulation, this expression is,
\begin{align}
    C_{AB}(t) = \int d\textbf{z}\left[\hat{\rho}\hat{A}\right](\textbf{z})\left[\hat{B}(t)\right](\textbf{z}), 
\end{align}
where $\left[.\right](\textbf{z})$ is the phase space function of the operator. 
In the following sections, for simplicity, we discuss the time dependence of 
expectation values (corresponding to $\hat{A}=\hat{1}$) noting that all the expressions derived can be applied 
to correlation functions by replacing $\hat{\rho}$ with $\hat{\rho}\hat{A}$.

The MQC correlation function is obtained by Filinov filtering
the phase of the quantum-limit DHK-IVR correlation function to 
obtain a phase space expression.~\cite{Antipov2015, Church2017}
Here we consider the MQC approximation to the 
time-dependent expectation value
of a general operator $\hat{B}$ for a 1D system 
\begin{align}
    \langle\hat{B}(t)\rangle_{\text{MQC}} &= \frac{1}{\left(2\pi\right)^{2}} \int d\textbf{z}_{0}\int d\textbf{z}_0^{\prime}\,D\left(\textbf{z}_{0},\textbf{z}_0^{\prime},c_{p},c_{q}\right) \notag \\ 
    & \times \mel{\textbf{z}_{0}}{\hat{\rho}}{\textbf{z}_{0}^{\prime}}  \mel{\textbf{z}_{t}^{\prime}}{\hat{B}}{\textbf{z}_{t}}  e^{i\left[S_t(\textbf{z}_{0})-S_{t}(\textbf{z}_{0}^{\prime})\right]}\notag \\ 
   &\times e^{-\frac{1}{2} c_{q} \Delta^{2}_{q_0}}\,e^{-\frac{1}{2} c_{p} {\Delta}^2_{p_0}},
   \label{cf-mqc-df}    
\end{align}
working in atomic units with $\hbar=1$ a.u.
In Eq.~\ref{cf-mqc-df}, $\hat{\rho}$ is the initial density operator for the system, 
$D\left(\textbf{z}_{0},\textbf{z}_0^{\prime},c_{p},c_{q}\right)$ is the 
MQC prefactor, $c_q$ and $c_p$ are 
Filinov parameters, and 
$\Delta_{x_0} = x_0 - x_0^{'}$ for $x \in \{p,q\}$
are the difference variables. 
The phase space variables, 
$\textbf{z}_t$ and $\textbf{z}_t^\prime$, in Eq.~\ref{cf-mqc-df} are obtained 
by propagating classical trajectories for time $t$ under the classical Hamiltonian
with initial conditions, $\textbf{z}_0,$ and $\textbf{z}_0^{\prime}$, respectively,
and $S_t(\textbf{z}_{0})$ and $S_{t}(\textbf{z}_{0}^{\prime})$ 
are the actions of forward and backward trajectories 
respectively. 

At time zero, the MQC prefactor is simply 
\begin{align}
    D = \sqrt{ (c_q+\gamma)({\gamma}^{-1}+c_{p})},
\end{align}
and the expectation value is, 
\begin{align}
    \langle\hat{B}(0)\rangle_{\text{MQC}} &= \frac{ \sqrt{(c_q+\gamma)({\gamma}^{-1}+c_{p})}}{\left(2\pi\right)^{2}} \int d\textbf{z}_{0}\int d\textbf{z}_0^{\prime} \mel{\textbf{z}_{0}}{\hat{\rho}}{\textbf{z}_{0}^{\prime}} \notag \\
   & \times \mel{\textbf{z}_{0}^{\prime}}{\hat{B}}{\textbf{z}_{0}}  e^{-\frac{1}{2} c_{q} \Delta^{2}_{q_0}}\,e^{-\frac{1}{2} c_{p} {\Delta}^2_{p_0}}.
   \label{cf-mqc-df-zerotime}    
\end{align}

For $\textbf{c}=\{c_p, c_q\}~=~0$, Eq.~\eqref{cf-mqc-df-zerotime} reduces to the DHK expression,
\begin{align}
    \langle\hat{B}(0)\rangle_{\text{DHK}} &= \lim_{\textbf{c}\to 0}\langle\hat{B}(0)\rangle_{\text{MQC}} \notag \\
    &=  \frac{1}{\left(2\pi\right)^{2}} \int d\textbf{z}_{0}\int d\textbf{z}_0^{\prime} \mel{\textbf{z}_{0}}{\hat{\rho}}{\textbf{z}_{0}^{\prime}}\mel{\textbf{z}_{0}^{\prime}}{\hat{B}}{\textbf{z}_{0}} \notag \\
    & = \text{Tr}\left[\hat{\rho}\hat{B}(0)\right]=\langle\hat{B}(0)\rangle_{\text{exact}},
   \label{cf-dhk-zerotime}   
\end{align}
where we use the coherent state completeness relation, 
\begin{align}
    \hat 1 =  \frac{1}{2\pi} \int d\textbf{z}_{0}\, \ket{\textbf{z}_{0}}\bra{\textbf{z}_{0}}.
\end{align}
In the classical limit (large Filinov parameters),  the MQC expectation value is 
\begin{align}
 \lim_{\textbf{c}\to \infty}\langle\hat{B}(0)\rangle_{\text{MQC}}  & = \frac{1}{2\pi} \int d\textbf{z}_{0}\,\mel{\textbf{z}_{0}}{\hat{\rho}}{\textbf{z}_{0}} \mel{\textbf{z}_{0}}{\hat{B}}{\textbf{z}_{0}} \notag \\
 &\neq \int d\textbf{z}_{0}\, \rho_H\left(\textbf{z}_{0}\right) \Tilde{B}_H\left(\textbf{z}_{0}\right) \notag \\
 & = \langle\hat{B}(0)\rangle_{\text{exact}},
 \label{b-cl-limit}
\end{align}
where the inequality follows from the definition of $\Tilde{B}_H$ in Eq.~\eqref{b-husimi}.

We have shown that while MQC is exact at time zero in the quantum limit ($\textbf{c} \to 0$), 
it is not, in general, exact in the classical limit ($\textbf{c} \to \infty $). 
To understand the behaviour of MQC at time zero for finite non-zero values of the Filinov
parameter, we insert the identity,~\cite{Pollak2022}
\begin{align}
    \hat{B} = \frac{1}{2\pi}\int d\textbf{z} \, \Tilde{B}_H(\textbf{z})\ketbra{\textbf{z}}{\textbf{z}}
\end{align}
into Eq.~\eqref{cf-mqc-df-zerotime} for both $\hat{\rho}$ and $\hat{B}$,
\begin{widetext}
\begin{align}
    \langle\hat{B}(0)\rangle_{\text{MQC}} &= \frac{ \sqrt{(c_q+\gamma)({\gamma}^{-1}+c_{p})}}{\left(2\pi\right)^{4}} 
    \int d\textbf{z}_{0}\int d\textbf{z}_0^{\prime} \int d\textbf{z}^{\prime}  \int d\textbf{z}^{\prime\prime} \braket{\textbf{z}_{0}}{\textbf{z}^{\prime}}
    \Tilde{\rho}_H\left(\textbf{z}^{\prime}\right)
    \braket{\textbf{z}^{\prime}}{\textbf{z}_{0}^{\prime}} 
    \braket{\textbf{z}_{0}^{\prime}}{\textbf{z}^{\prime\prime}}\Tilde{B}_H(\textbf{z}^{\prime\prime})
    \braket{\textbf{z}^{\prime\prime}}{\textbf{z}_{0}} 
    e^{-\frac{1}{2} c_{q} \Delta^{2}_{q_0}}
    \,e^{-\frac{1}{2} c_{p} {\Delta}^2_{p_0}} \label{mqc-t0-1}\\
    & = \int d\textbf{z} \, \rho_H(\textbf{z})
    \left[ \hat{\mathcal{G}}\left(\Tilde{\gamma}_q;q\right) \hat{\mathcal{G}}\left(\Tilde{\gamma}_p^{-1};p\right)\Tilde{B}_H(\textbf{z}) \right] \label{mqc-t0-4} \\
    &\neq \int d\textbf{z} \, \rho_H\left(\textbf{z}\right) \Tilde{B}_H\left(\textbf{z}\right) = \langle\hat{B}(0)\rangle_{\text{exact}}, \label{mqc-t0-5} 
\end{align}
\end{widetext}
where $\gamma$ in Eq.~\eqref{mqc-t0-1} is the coherent state width, 
and $\Tilde{\gamma}_q$ and $\Tilde{\gamma}_p$ in Eq.~\eqref{mqc-t0-4} are functions of $\gamma$ and the 
Filinov parameters, $c_q$ and $c_p$, respectively. 
Detailed definitions of $\Tilde{\gamma}_q$ and $\Tilde{\gamma}_p$ are 
provided in Appendix ~\ref{app:derivation} along with 
details of the intermediate steps in the derivation.\\
The inequality in Eq.~\eqref{mqc-t0-4} indicates that for finite values
of the Filinov parameter, MQC is not exact at zero time for a general operator $\hat B$.
The exceptions are the position and momentum operators, 
$\hat{B} \equiv \hat{x}$ or $\hat{p}$, where 
\begin{align}
\hat{\mathcal{G}}\left(\Tilde{\gamma}_q;q\right) \hat{\mathcal{G}}\left(\Tilde{\gamma}_p^{-1};p\right)\Tilde{B}_H(\textbf{z}) = \Tilde{B}_H(\textbf{z}),
\end{align}
for all values of {\bf c}, a
finding that is consistent with previously published numerical results
for correlation functions involving these operators.~\cite{Antipov2015,Church2017}

We have shown that for a general operator, $\hat B$, the MQC correlation
function is inaccurate at time zero for non-zero values of the 
Filinov parameter, and that its classical-limit does not reproduce
the Husimi-IVR expression. One potential approach to this problem
would be to reformulate MQC by modifying the choice of phase that 
is filtered. In the next section, we describe a second strategy that 
allows us to work within the present MQC formalism using
a modified operator $\hat B$.

\subsection{Modifying the MQC Correlation Function}

Comparing Eq.~\eqref{mqc-t0-4} and Eq.~\eqref{mqc-t0-5} 
suggests that we can correct the MQC correlation function
by replacing operator $\hat B$ by a new operator $\hat{\mathcal{B}}$, such that
\begin{align}
    \hat{\mathcal{G}}\left(\Tilde{\gamma}_q;q\right) \hat{\mathcal{G}}\left(\Tilde{\gamma}_p^{-1};p\right)\Tilde{\mathcal{B}}_H(\textbf{z}) = \Tilde{B}_H(\textbf{z}).
    \label{mod-b-def-h}
\end{align}
or equivalently using the relationship between the Anti-Husimi transform
and the Wigner transform in Eq.~\eqref{anti-hus-def},
\begin{align}
    \hat{\mathcal{G}}\left(\Tilde{\gamma}_q;q\right) \hat{\mathcal{G}}\left(\Tilde{\gamma}_p^{-1};p\right)\mathcal{B}_W(\textbf{z}) = B_W(\textbf{z}).
\end{align}
The MQC expression with the modified operator is 
still exact in the quantum limit;
when $\textbf{c}\to 0 $,
the Gaussian convolutions in  Eq.~\eqref{mod-b-def-h} reduce to delta functions, 
such that 
\begin{align}
    \hat{\mathcal{B}} = \hat{B}.
\end{align}
In the classical limit, $\textbf{c}\to \infty$, the Filinov-dependent 
width parameters $\Tilde{\gamma_q} \to \frac{\gamma}{2}$ and $\Tilde{\gamma_p} \to 2\gamma$,
and using Eq.~\eqref{hus-anti-hus} we obtain 
\begin{align}
\lim_{\textbf{c}\to\infty}\mathcal{B}_H(\textbf{z}) = \Tilde{B}_H(\textbf{z}).
\end{align}
The modified MQC expression thus achieves the correct form in 
both the quantum and classical limits, with
\begin{align}
     \lim_{\textbf{c}\to 0}\langle\hat{\mathcal{B}}(t)\rangle_{\text{MQC}} &
     = \langle\hat{B}(t)\rangle_{\text{DHK}},
\end{align}
and
\begin{align}
     \lim_{\textbf{c}\to \infty}\langle\hat{\mathcal{B}}(t)\rangle_{\text{MQC}} &
     = \langle\hat{B}(t)\rangle_{\text{Hus}}.
 \end{align}

Although our analysis and derivation of the modified operator
is shown here for the expectation value, this idea can 
be simply extended to real-time correlation functions 
resulting in the modified MQC expression,
\begin{align}
    C_{AB}(t) &=
    \frac{1}{\left(2\pi\right)^{2}} \int d\textbf{z}_{0}\int d\textbf{z}_0^{\prime}\,D\left(\textbf{z}_{0},\textbf{z}_0^{\prime},c_{p},c_{q}\right) \notag \\ 
    & \times \mel{\textbf{z}_{0}}{\hat{\rho}\hat{A}}{\textbf{z}_{0}^{\prime}}  \mel{\textbf{z}_{t}^{\prime}}{\hat{\mathcal{B}}}{\textbf{z}_{t}}  e^{i\left[S_t(\textbf{z}_{0})-S_{t}(\textbf{z}_{0}^{\prime})\right]}\notag \\ 
   &\times e^{-\frac{1}{2} c_{q} \Delta^{2}_{q_0}}\,e^{-\frac{1}{2} c_{p} {\Delta}^2_{p_0}}.
   \label{mqc-df-mod}    
\end{align}
Finding the modified operator $\hat{\mathcal{B}}$ can 
be non-trivial, but for simple operators it is possible to 
derive an analytic expression. 
For instance, the modified MQC correlation function 
when $\hat B~=~\hat x^2$ is obtained with 
the modified operator $\hat \chi^2$, 
\begin{align}
    \hat x^2 \to \hat{\chi}^2=\hat{x}^2 - \frac{c_p}{1+\gamma c_{p}},
    \label{mod-x2}
\end{align}
and similarly the modified MQC correlation function for 
$\hat B~=~\hat p^2$ is obtained with the modified operator, $\hat \pi^2$,
\begin{align}
    \hat{p}^2 \to \hat{\pi}^2 = \hat{p}^2 - \frac{\gamma c_{q}}{\gamma +c_{q}}.
    \label{mod-p2}
\end{align}

\section{Model System and Simulation Details}\label{sec:simulation}
Inspired by a model previously used to study 
ZPE leakage in SC dynamics,~\cite{Buchholz2018} we consider three 
different models of coupled harmonic oscillators with cubic 
couplings,
\begin{align}
    V(x_1,..,x_F) = \sum_{i=1}^{F} \frac{1}{2}m\omega_{i}^2x_{i}^2 + \sum_{i<j} C_{ij}\left(x_i-x_j\right)^3,
\end{align}
where F is number of degrees of freedom, all oscillators have mass $m = 1$ a.u., 
$\omega_i$ is the frequency of the i$^{th}$ oscillator, 
and $C_{ij}$ is the pair-wise coupling between oscillators. 
Model A is a two dimensional model \textemdash\, two harmonic oscillators 
with cubic coupling. Models B and C are three dimensional models where 
one oscillator is weakly coupled to a subsystem comprising two 
strongly coupled oscillators. 
In model C, two weakly coupled oscillators have the same
frequency to explore the effect of having resonant modes 
treated with different levels of quantization.
Frequencies of the oscillators and coupling constants for all models are listed in Table~\ref{tbl:model-parameters}.

\begin{table}
    \caption{Parameters for the three model systems (in atomic units).}
  \label{tbl:model-parameters}
  \begin{ruledtabular}
  \begin{tabular}{lccccccc}
    Model   &  F    & $\omega_1$  & $\omega_2$ &  $\omega_3$ & $C_{12}$ & $C_{23}$ & $C_{13}$ \\
    \hline
    A       & 2     & 1.0           & 0.5 & - & $10^{-4}$ & - & - \\
    B       & 3     & 1.0           & 0.5 & 0.25 & $10^{-4}$ & $2 \times 10^{-4}$ & 0.0 \\
    C       & 3     & 1.0           & 1.0 & 0.5 & $10^{-4}$ & $2 \times 10^{-4}$ & 0.0 \\
  \end{tabular}
  \end{ruledtabular}
\end{table}

We chose the initial coherent state $|\psi_i\rangle \equiv |{\bf z}_i \rangle$ 
to correspond to the ground-state of the uncoupled harmonic oscillators, 
a product of coherent states with widths $\gamma_i = m\omega_i$, 
centered at ${\bf z}_i=(0,0)$. Since the magnitude of the coupling 
constants is very small, the overlap between the initial state and the ground state 
of the coupled system is nearly unity ensuring that only the lowest
vibrational state for each oscillator is populated. The total energy can
then be written as,
\begin{align}
    E=\sum_{i} E_i 
\end{align}
where the $E_i$ is the ZPE of the $i^\text{th}$ oscillator.~\cite{Buchholz2018}

To track the flow of ZPE, we calculate the energy expectation value 
of each oscillator as a function of time,
\begin{align}
    \langle \hat{E}_i(t) \rangle = \frac{1}{2m} \langle \hat{p}_{i}^{2}(t) \rangle 
        + \frac{1}{2} m \omega_{i}^{2} \langle \hat{x}_{i}^{2}(t) \rangle. 
        \label{eq:eiexpect}
\end{align}
In our modified MQC implementation, the energy expectation value in Eq.~\ref{eq:eiexpect}
is calculated by evaluating two independent non-linear correlation functions
using Eq.~\eqref{mqc-df-mod}. Specifically, 
the $\left\langle \hat x_i^2(t)\right\rangle$ term is obtained using 
modified operator $\hat{\mathcal{B}}=\hat \chi_i^2(t)$ defined in Eq.~\eqref{mod-x2} 
and the $\left\langle \hat p_i^2(t)\right\rangle$ term is obtained 
using modified operator $\hat{\mathcal{B}}=\hat \pi_i^2(t)$ defined in Eq.~\eqref{mod-p2}.
Numerical integrals are evaluated using standard Monte Carlo techniques with the sampling function,
\begin{align}
    \rho\left(\textbf{z}_0,\textbf{z}_0^{\prime}\right) 
    &= \mathcal{N} \, \lvert\braket{{\textbf{z}}_{0}}{\psi_i}\braket*{\psi_i}{{\textbf{z}}_{0}^{\prime}}\rvert e^{-\frac{1}{2}\left({\bf q}_0-{\bf q}_0^{\prime}\right)^{T}.\textbf{c}_{q}.\left({\bf q}_0-{\bf q}_0^{\prime}\right)} \notag \\
    & \times  e^{-\frac{1}{2}\left({\bf p}_0-{\bf p}_0^{\prime}\right)^{T}.\textbf{c}_{p}.\left({\bf p}_0-{\bf p}_0^{\prime}\right)},
\end{align}
where $\mathcal{N}$ ensures proper normalization. 
Note that while the sampling is performed using sum and difference
variables, we then transform back into the 
$(\textbf{z}_0,\textbf{z}_0^{'})$ for trajectory propagation 
and calculation of the estimator. Trajectories are 
propogated under the classical Hamiltonian using a fourth-order symplectic integrator\cite{Brewer1997}
with a timestep of 0.1 a.u. for Model A, and 0.175 a.u. for models B and C to ensure total energy
conservation. All calculations presented here were converged
to within the error bars shown using at most $10^7$  trajectories.

Finally, we note that we renormalize the energy in
each mode such that the total energy stays constant 
at all times, $\sum_{i=1}^{F} \langle \hat{E}_{i}(t) 
\rangle = \sum_{i=1}^{F} \langle \hat{E}_{i}(0) \rangle $. 
This is required because SC methods used here are, at best,
only approximately unitary.\cite{Herman1986,Garashchuk1997,Harabati2004,Zhang2005,Tatchen2011}

\section{Results and Discussion} \label{sec:results}

Using similar model systems, previous work has established that 
DHK-IVR, a quantum-limit SC method, conserves ZPE whereas LSC-IVR, 
a classical-limit SC method does not.~\cite{Buchholz2018} 
Here, we use the modified MQC method to establish the extent of ZPE
leakage in dynamics that employ finite values of the Filinov parameter. 
Further, we investigate how selective quantization of modes influences
ZPE flow, with the potential to limit the extent to which 
the unphysical flow of energy occurs within a subsystem.

\begin{figure}[h]
    \centering
    \includegraphics[width=0.45\textwidth]{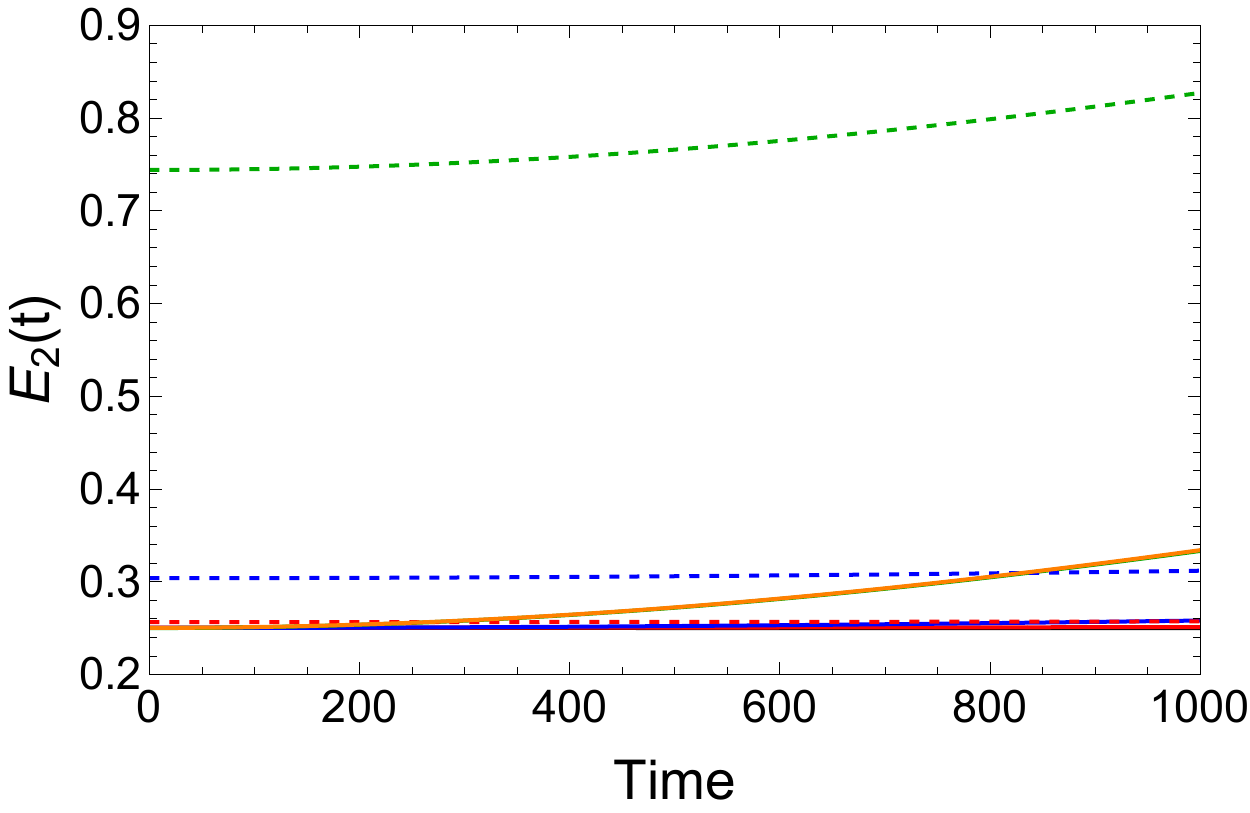}
    \caption{ We plot the energy expectation value for the low frequency
    oscillator 2 as a function of time for model A. 
    The original MQC results are shown using dashed lines and compared against 
    the modified MQC method shown using solid lines for different values of 
     the Filinov parameter $c = 0.01$ (red), $c = 0.1$ (blue), and $c =100$ (green). 
      Results from DHK-IVR (black) and Husimi-IVR (orange) are also shown for reference. 
        Note that in the quantum-limit, $c = 0.01$ both the original and modified 
        MQC results coincide with the DHK-IVR result,
        whereas in the classical limit, $c=100$, only the modified MQC reproduces 
        the Husimi-IVR results with the original formulation yielding results that 
        are significantly different even at time zero.
    }
    \label{fig:mqc_oldvnew}
\end{figure}

In Fig.~\ref{fig:mqc_oldvnew}, we demonstrate that the modified MQC expression,
unlike the original, reproduces the exact values for
the energy expectation value of a single oscillator in 2D model A 
for all values of the Filinov parameter.
We also show that the classical limit of the modified MQC also correctly
reproduces the Husimi IVR result with significant ZPE leakage.

\begin{figure}[h]
    \centering
    \includegraphics[width=0.45\textwidth]{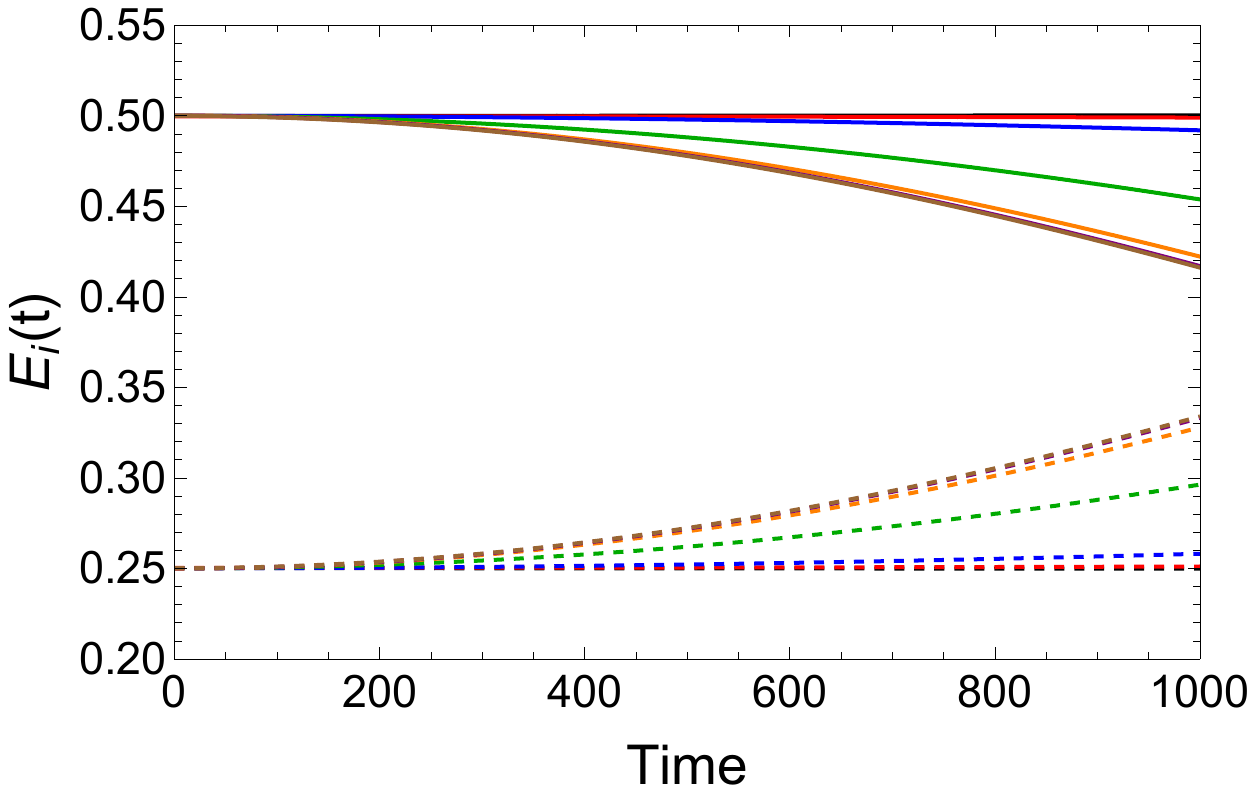}
    \caption{We plot the energies of high frequency oscillator 1 (solid lines) and low frequency oscillator 2 (dashed lines) as a function of time for Model A calculated using DHK-IVR (black), MQC with $\textbf{c} = 0.01$ (red), $\textbf{c} = 0.1$ (blue), $\textbf{c} = 1$ (green), $\textbf{c} = 10$ (orange), $\textbf{c} = 100$ (purple), and Husimi-IVR (brown). DHK-IVR results overlap with $\textbf{c} = 0.01$, and Husimi-IVR results overlap with $\textbf{c} = 100$. As expected, the extent of ZPE leakage increases systematically as the value of Filinov parameters is increased}
    \label{2d-ceq}
\end{figure}

Having established the accuracy of the modified MQC expression in capturing
both quantum-limit and classical-limit SC dynamics, 
we explore the extent to which this method conserves ZPE for different 
values of the Filinov parameter. In Fig.~\ref{2d-ceq} we plot the 
average energies of the two oscillators in model A. 
As expected, DHK-IVR and quantum-limit  MQC,  
$\textbf{c}_1=\textbf{c}_2=\textbf{c} = 0.01$, 
both show ZPE conservation for the length of the simulation. 
On the other hand, Husimi-IVR and classical-limit MQC, $(\textbf{c} = 100)$, 
both exhibit significant loss of ZPE with unphysical energy flow from the 
high frequency oscillator into the low frequency one. 
As the Filinov parameter is increased, we 
see a systematic onset of ZPE leakage from the high frequency mode,
consistent with the idea that SC methods rely on interference to conserve 
ZPE. It is notable that despite increasing the Filinov parameter 10-fold,
MQC with $c=0.1$ still exhibits reasonable ZPE conservation and requires
far few trajectories for convergence than the corresponding $c=0.01$ MQC 
simulation. 

\begin{figure}[h]
    \centering
    \includegraphics[width=0.45\textwidth]{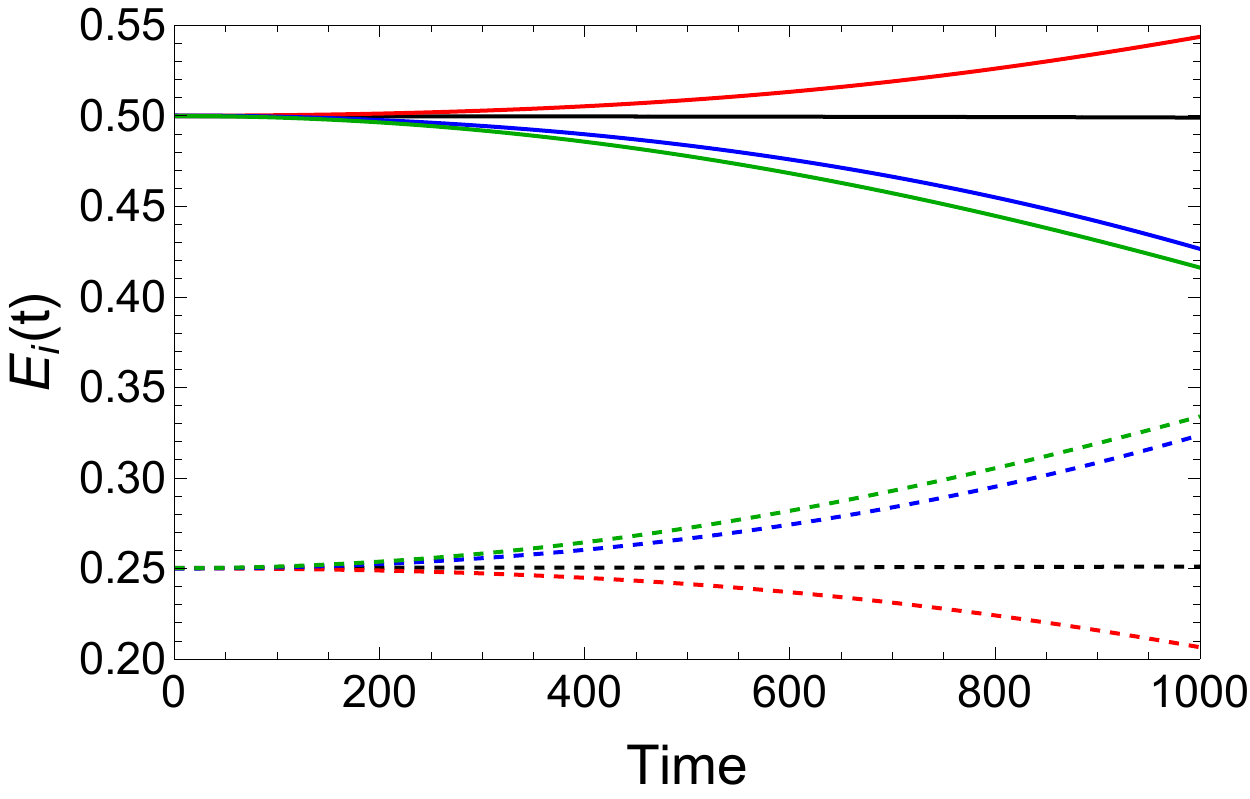}
    \caption{We plot the individual oscillator energies obtained 
        when performing a mixed quantum-classical MQC simulation 
        with the energy of the high frequency mode (oscillator 1) shown using 
        solid lines and the low frequency mode (oscillator 2) shown with dashed lines. Results obtained by treating all modes 
        in the quantum limit with $\textbf{c}_1 = \textbf{c}_2 = 0.01$ are
        shown in black and compared against two mixed-limit cases with $\textbf{c}_1 = 0.01, \textbf{c}_2 = 100$ (red), 
    and $\textbf{c}_1 = 100, \textbf{c}_2 = 0.01$ (blue), 
    as well as the classical limit $\textbf{c}_1 = \textbf{c}_2 = 100$ (green).}
    \label{2d-mixed}
\end{figure}

Working with 2D model A, we also perform MQC simulations where one mode 
is treated in the quantum limit while the other is described 
in the classical limit. In Fig.~\ref{2d-mixed}, we demonstrate 
that the mode that is treated in the quantum limit does indeed 
conserve ZPE, whereas the classical-limit mode does not. Interestingly,
when we quantize the low frequency oscillator, there is no observed change
in ZPE flow with the classical-limit high-frequency mode continuing 
to lose energy. However, when we treat the high frequency mode in the 
quantum limit, we stem the loss of energy from this mode and instead 
see a reversal with energy flowing uphill from the classical-limit
low-frequency mode.

\begin{figure}[h]
    \centering
    \includegraphics[width=0.45\textwidth]{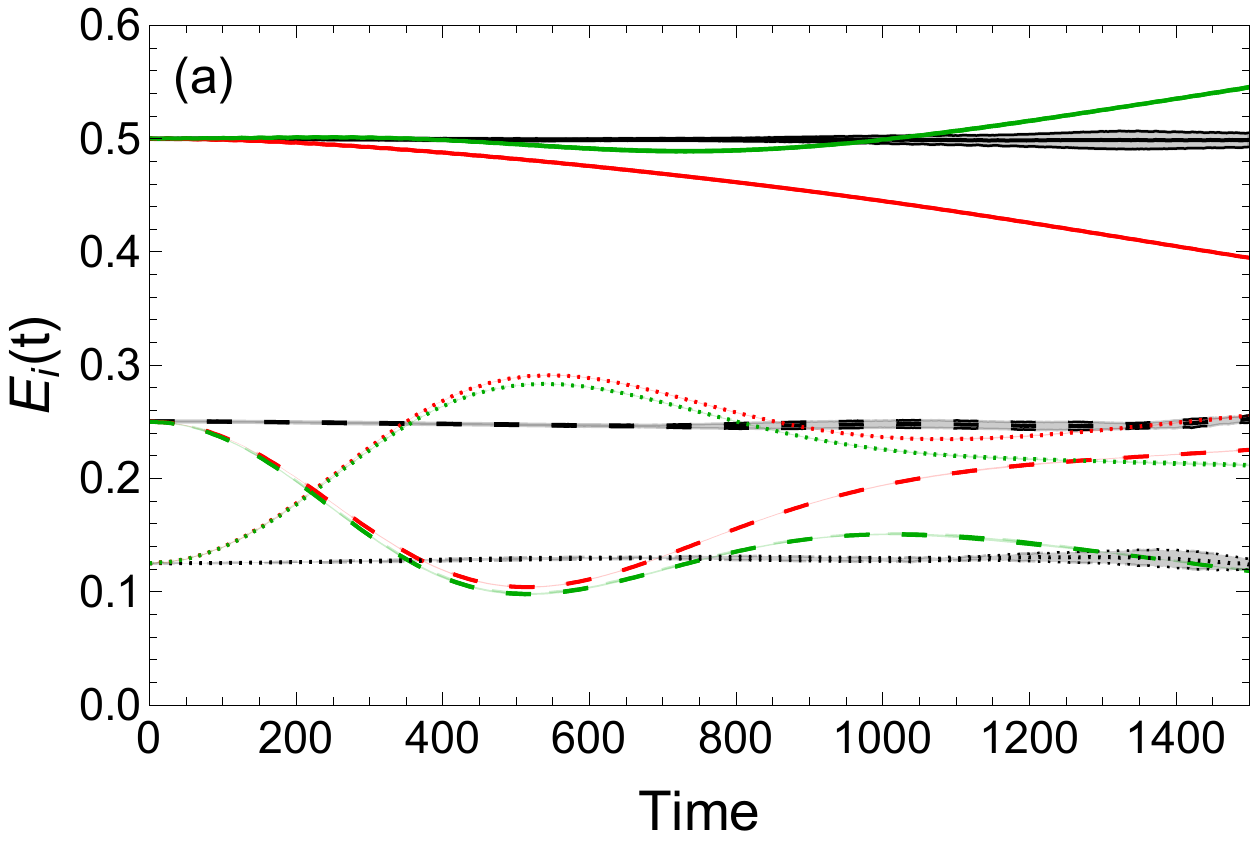}
    \includegraphics[width=0.45\textwidth]{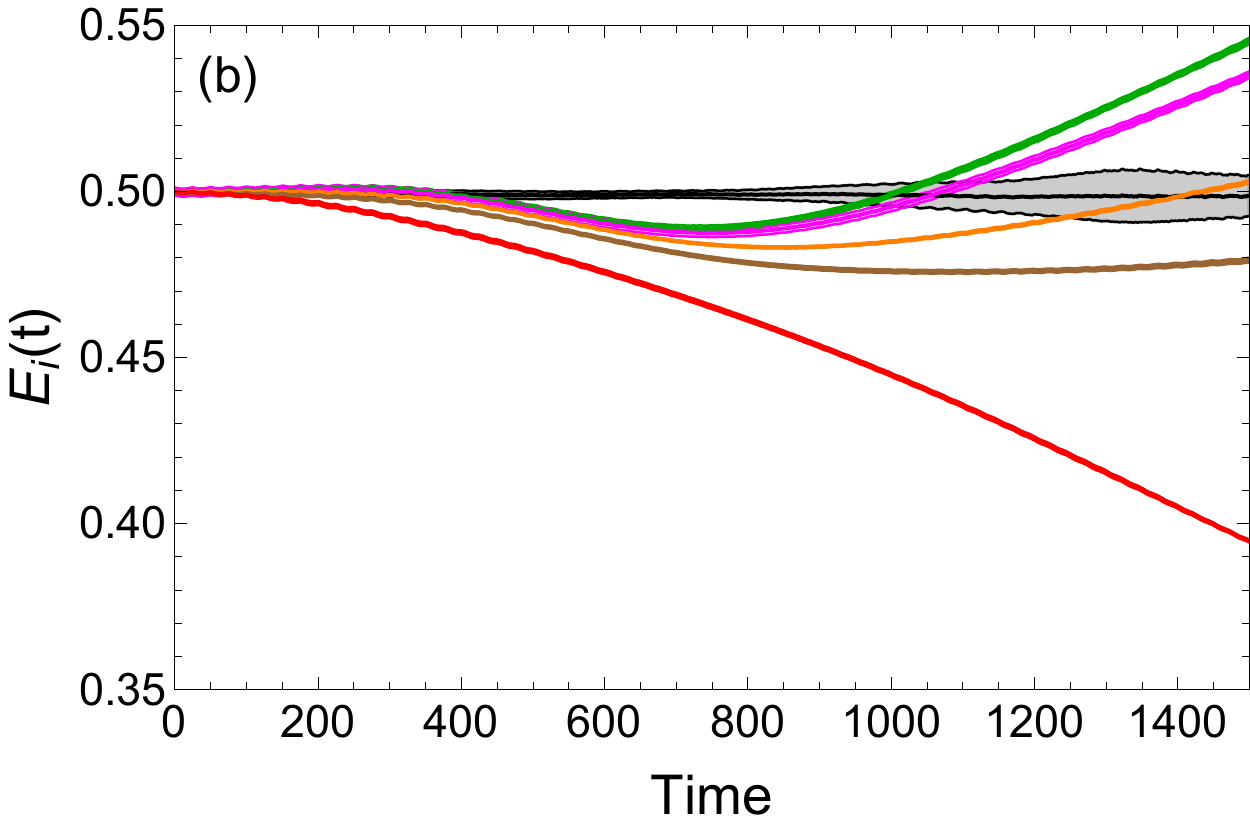}
    \includegraphics[width=0.45\textwidth]{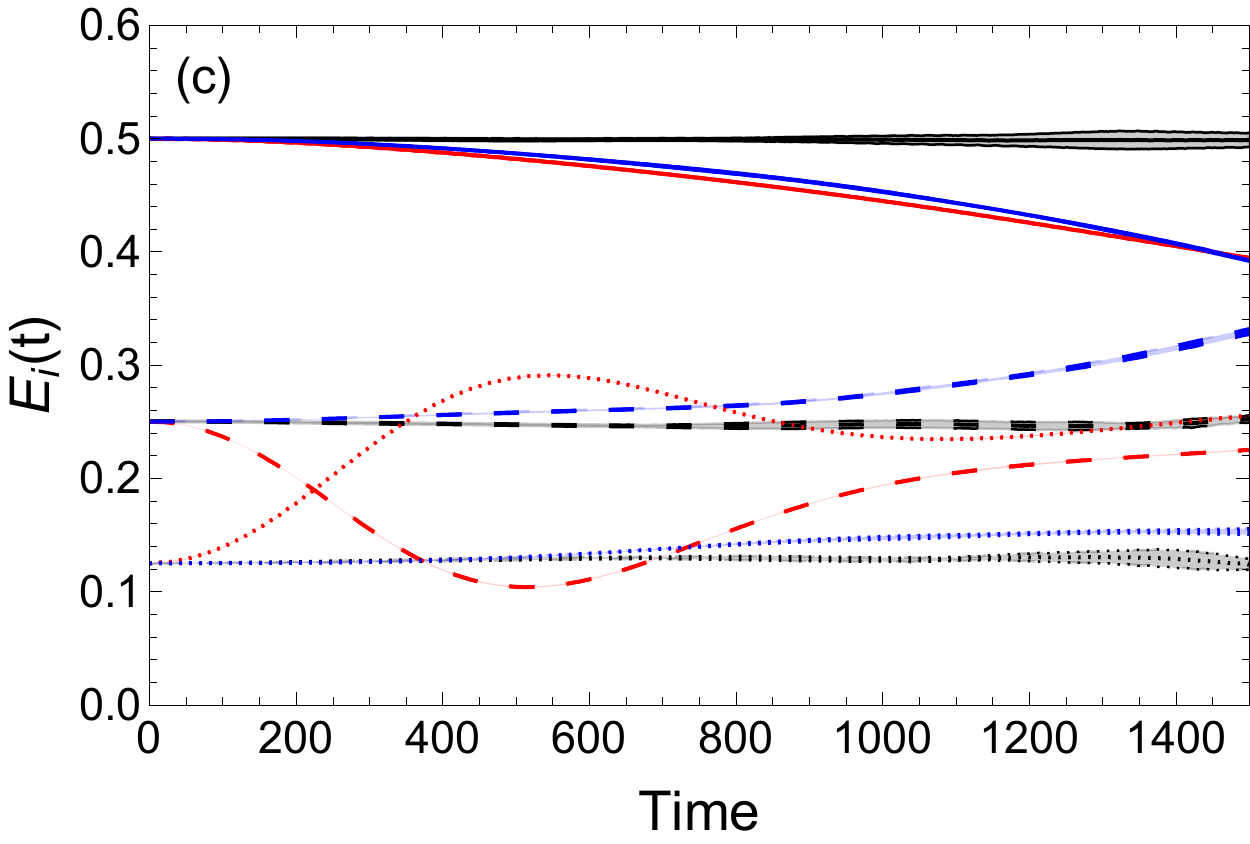}
    \caption{We plot the energy as a function of time in oscillators 1 (solid), 2 (dashed) and 3 (dotted) in model B. MQC results in the quantum-limit ($\textbf{c}=0.01$, black) and classical-limit ($\textbf{c}=1000$, red) are shown in all panels and compared with different versions of mixed quantization. (a) weakly coupled high frequency oscillator 1 is treated in the quantum limit, while the other two strongly coupled lower frequency modes are treated classically, $\textbf{c}_1 = 10^{-2}, \textbf{c}_2=\textbf{c}_3=10^3$ (green). (b) Oscillators 2 and 3 are treated classically $(\textbf{c}_2 = \textbf{c}_3 = 10^3)$ again, and the Filinov parameter for oscillator 1 is varied  \textemdash $\textbf{c}_1 = 10^{-2}$ (green), $\textbf{c}_1 = 0.1$ (pink), $\textbf{c}_1 = 0.5$ (orange), and $\textbf{c}_1 = 1$ (brown). (c) The two lower frequency oscillators are treated in the quantum-limit eliminating the fast ZPE exchange observed in the classical-limit simulations with 
    $\textbf{c}_1 = 10^{3}, \textbf{c}_2=\textbf{c}_3=10^{-2}$ (blue). In all cases, the quantized sub-system conserves ZPE for a significantly longer time compared to the classical-limit, before eventually accepting energy from the classical sub-system. Highlights around lines indicate error bars.}
    \label{modelb-plots}
\end{figure}

Model B offers a more interesting test case for mixed quantization. 
For this model, in the quantum limit, all three oscillators conserve 
ZPE as shown in Fig.~\ref{modelb-plots}, and 
in the classical limit, we see rapid ZPE exchange between the two
lower frequency oscillators that are strongly coupled, 
with the weakly coupled high frequency oscillator losing 
ZPE to the other two modes on a longer timescale.
In the first study of mixed quantization, we treat
the high frequency oscillator in the quantum limit
while describing the rest in the classical limit.
As shown in Fig.~\ref{modelb-plots}(a), we find that the classical `bath' 
oscillators continue to exchange ZPE but the quantum-limit oscillator
approximately conserves ZPE over a significantly longer timescale 
(by a factor of 2) with some increase in energy corresponding to leakage 
from the classical subsystem.

\FloatBarrier

\begin{figure}[h!]
    \centering
    \includegraphics[width=0.45\textwidth]{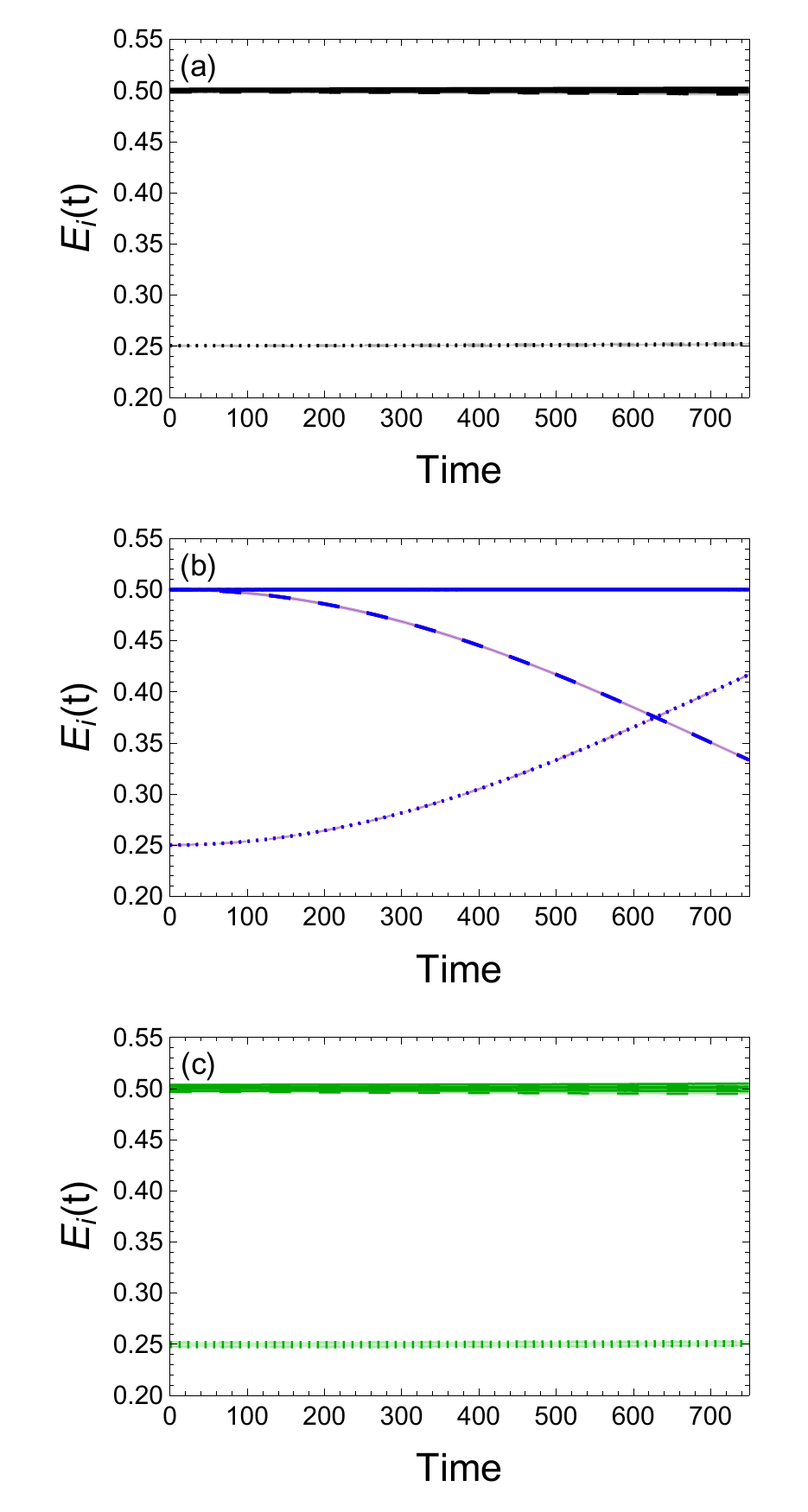}
    \caption{We plot the energy as a function of time for high frequency oscillators 1 (solid) and 2 (dashed), and low frequency oscillator 3 (dotted) in model C, compared for various different MQC limits \textemdash  (a) All modes treated in the quantum limit with $\textbf{c}=0.01$ (black); (b) All modes treated in the classical limit with $\textbf{c}=100$ (red) yield results that are identical to the mixed 
    quantization case where just the weakly coupled oscillator 1 is treated in the quantum limit, $\textbf{c}_1 = 0.01, \textbf{c}_2 = \textbf{c}_3 = 100$ (blue), and (c) We quantize the strongly coupled subsystem of oscillator 2 and 3 with the resonant, high frequency oscillator 1 treated in the classical limit, $\textbf{c}_1 = 100, \textbf{c}_2 = \textbf{c}_3 = 0.01$ (green). Note that in (a) and (c), solid and dashed lines overlap. In (b), results for the two different Filinov parameter values (red and blue) overlap. Highlights around lines indicate error bars.}
    \label{modelc-plots}
\end{figure}

In a final study of model B, 
we quantize the strongly coupled subsystem of low frequency oscillators and treat the high frequency oscillator in the classical limit,
and plot the results in Fig.~\ref{modelb-plots}(c)
We find that the quantized subsystem now conserves ZPE
with no exchange of energy between the two low frequency 
oscillators. There is a net increase in the individual oscillator 
energies of the subsystem, however, due to energy flowing 
into the quantized subsystem from the classical-limit, 
high frequency oscillator.

Model C is similar to model B, but explores how the presence
of oscillators of equal frequency (resonant) in the classical
and quantum-limit subsystems affects overall ZPE flow. 
Specifically, we have two weakly coupled high frequency oscillators, 
one of which is strongly coupled to the low frequency oscillator.
If all three are treated in the quantum limit, the expected 
ZPE conservation is seen in the MQC results in Fig.~\ref{modelc-plots}.
Interestingly, when all modes are treated in the classical limit, we 
see some exchange of ZPE between the two strongly coupled oscillators,
but no ZPE leakage from the weakly coupled oscillator perhaps 
as a consequence of the resonance in the model. 
Given this, quantizing just the weakly coupled oscillator continues to conserve its energy while the classical limit oscillators
exhibit ZPE exchange. 
However, quantizing the strongly coupled oscillators, results 
in conservation of ZPE - a promising insight into conservation strategies 
for systems where the quantized subsystem and classical bath 
share a resonant frequency.

\section{Conclusions} \label{sec:conclusion}

In this paper, we begin with a detailed study of the origin 
of the MQC SC method's inaccuracy at time zero for nonlinear
correlation functions. We show that MQC with finite, non-zero
values of the Filinov parameter is not exact at time zero
for a general operator $\hat B$. Further, we show that 
the real-time MQC correlation functions in the classical limit
does not to the Husimi IVR correlation function. 
We perform a detailed analysis of the Husimi and Wigner phase
space formulation of quantum mechanics and identify a simple 
modification to correct for both these problems. 
Specifically, we propose that replacing operator $\hat B$ with a 
new operator that is obtained through inverse Gaussian 
convolutions of Wigner/Anti-Husimi functions of 
the original operator. We analytically derive the modified 
operator for $\hat x^2$ and $\hat p^2$, and use the resulting 
expression to characterize ZPE flow in MQC simulations. 

We construct a series of model systems comprising oscillators
of different frequencies and with different coupling motifs 
chosen to mimic real system connectivities. We demonstrate
that by selective quantization of weakly coupled modes, it is 
possible to improve the timescale on which ZPE is conserved for 
a subset of modes. This work establishes a promising path forward
to real system simulations using the modified MQC approach proposed
here. Next steps include establishing a path to deriving modified
operators corresponding to other operators, including most notably
the projection operator.

\begin{acknowledgments}
    This work was funded, in part, by NSF CAREER Grant No. CHE 1555205.
    The authors thank Jennifer Mavroudakis for helpful discussions.
\end{acknowledgments}

\appendix

\begin{widetext}
\section{Detailed derivation of the modified MQC method}\label{app:derivation}
Starting with Eq.\eqref{mqc-t0-1}, we note that the integrals over $\textbf{z}_0$ and $\textbf{z}_0^{\prime}$ can be evaluated analytically to yield,
\begin{align}
    \langle\hat{B}(0)\rangle_{\text{MQC}} &= \frac{ \sqrt{(c_q+\gamma)({\gamma}^{-1}+c_{p})}}{\left(2\pi\right)^{4}} 
    \int d\textbf{z}_{0}\int d\textbf{z}_0^{\prime} \int d\textbf{z}^{\prime}  \int d\textbf{z}^{\prime\prime} \braket{\textbf{z}_{0}}{\textbf{z}^{\prime}}
    \Tilde{\rho}_H\left(\textbf{z}^{\prime}\right)
    \braket{\textbf{z}^{\prime}}{\textbf{z}_{0}^{\prime}} 
    \braket{\textbf{z}_{0}^{\prime}}{\textbf{z}^{\prime\prime}}\Tilde{B}_H(\textbf{z}^{\prime\prime})
    \braket{\textbf{z}^{\prime\prime}}{\textbf{z}_{0}} 
    e^{-\frac{1}{2} c_{q} \Delta^{2}_{q_0}}
    \,e^{-\frac{1}{2} c_{p} {\Delta}^2_{p_0}} \notag \\ 
    & = \frac{1}{2\pi} \sqrt{\frac{\gamma_q}{\pi}}\sqrt{\frac{1}{\gamma_p\pi}} \int d\textbf{z}^{\prime}\, \int d\textbf{z}^{\prime\prime} e^{-\gamma_q\left(q^{\prime}-q^{\prime\prime}\right)^2}\,e^{-\gamma_p^{-1}\left(p^{\prime}-p^{\prime\prime}\right)^2}    \Tilde{\rho}_H\left(\textbf{z}^{\prime}\right)
    \Tilde{B}_H(\textbf{z}^{\prime\prime})
    \label{mqc-t0-a2}
\end{align}
where $\gamma_{q} = \frac{\gamma}{2}\frac{\gamma^{-1}+c_p}{\gamma^{-1}+2c_p}$, and $\gamma_{p}^{-1} = \frac{1}{2\gamma}\frac{c_q+\gamma}{2c_q+\gamma}$. We then split the Gaussian convolution with width $\gamma_3$ into two separate Gaussian convolutions with widths $\gamma_1$ and $\gamma_2$,
\begin{align}
    \sqrt{\frac{\gamma_3}{\pi}} \int dx^{\prime} e^{-\gamma_3\left(x-x^{\prime}\right)^2} f\left(x^{\prime}\right) = \sqrt{\frac{\gamma_1}{\pi}} \int dx^{\prime\prime} e^{-\gamma_1\left(x-x^{\prime\prime}\right)^2} \left[\sqrt{\frac{\gamma_2}{\pi}} \int dx^{\prime} e^{-\gamma_2\left(x^{\prime\prime}-x^{\prime}\right)^2}f\left(x^{\prime}\right) \right],
    \label{combine-gc}
\end{align}
such that $\gamma_3^{-1}=\gamma_1^{-1}+\gamma_2^{-1}$. We now define $\Tilde{\gamma}_{q} = \frac{\gamma}{2}\frac{\gamma^{-1}+c_p}{c_p}$, and $\Tilde{\gamma}_{p}^{-1} = \frac{1}{2\gamma}\frac{c_q+\gamma}{c_q}$, such that
\begin{align}
   &  \gamma_q^{-1} = \Tilde{\gamma}_q^{-1} + 2\gamma^{-1}, 
\end{align}
and 
\begin{align}
    & \gamma_p = \Tilde{\gamma}_p + 2\gamma,
\end{align}
to split the Gaussian convolutions in Eq.\eqref{mqc-t0-a2} into two Gaussian convolutions each,
\allowdisplaybreaks
\begin{align}
    \langle\hat{B}(0)\rangle_{\text{MQC}} &= \frac{1}{2\pi} \int d\textbf{z} \left[\frac{1}{2\pi} \int d\textbf{z}^{\prime} e^{-\frac{\gamma}{2}\left(q-q^{\prime}\right)^2}\,e^{-\frac{1}{2\gamma}\left(p-p^{\prime}\right)^2} \Tilde{\rho}_H\left(\textbf{z}^{\prime}\right)  \right]\left[\sqrt{\frac{\Tilde{\gamma}_q}{\pi}}\sqrt{\frac{1}{\Tilde{\gamma}_p\pi}}  \int d\textbf{z}^{\prime\prime} e^{-\Tilde{\gamma}_q\left(q-q^{\prime\prime}\right)^2}\,e^{-\Tilde{\gamma}_p^{-1}\left(p-p^{\prime\prime}\right)^2} \Tilde{B}_H(\textbf{z}^{\prime\prime}) \right]
     \label{mqc-t0-a3} \\
     & = \frac{1}{2\pi} \int d\textbf{z} \,\left[\hat{\mathcal{G}} \left(\frac{\gamma}{2};q\right) \hat{\mathcal{G}}\left(\frac{1}{2\gamma};p\right) \Tilde{\rho}_H\left(\textbf{z}\right) \right] \times \left[ \hat{\mathcal{G}}
   \left(\Tilde{\gamma}_q;q\right) \hat{\mathcal{G}}\left(\Tilde{\gamma}_p^{-1};p\right)\Tilde{B}_H(\textbf{z}) \right] \label{mqc-t0-a4} \\ 
   & = \int d\textbf{z} \, \rho_H(\textbf{z})
    \left[ \hat{\mathcal{G}}\left(\Tilde{\gamma}_q;q\right) \hat{\mathcal{G}}\left(\Tilde{\gamma}_p^{-1};p\right)\Tilde{B}_H(\textbf{z}) \right]. \label{mqc-t0-a5}
\end{align}
To go from Eq.\eqref{mqc-t0-a3} to Eq.\eqref{mqc-t0-a4}, we have used the definition of a Gaussian convolution from Eq.\eqref{gc-1}, and then used the relationship between the Husimi transform and the anti-Husimi transform of an operator, Eq.\eqref{hus-anti-hus}, to yield Eq.\eqref{mqc-t0-a5}, which is the same as Eq.\eqref{mqc-t0-4}.
\end{widetext}

\bibliographystyle{apsrev4-2}

\bibliography{bibfile}

\end{document}